\documentclass[twocolumn,preprintnumbers]{revtex4}

\usepackage[dvipdfmx]{graphicx}
\usepackage{epsf}
\usepackage{color}
\usepackage{amsmath,amssymb}
\usepackage{ascmac}
\usepackage{bm}
\usepackage{color}
\usepackage{ulem}

\usepackage{version}	
\begin{document}

\title{Lattice QCD study of static quark and antiquark correlations via entanglement entropies}

\author{Toru T. Takahashi}
\affiliation{National Institute of Technology, Gunma college, Maebashi, Gunma
371-8530, Japan}
\author{Yoshiko Kanada-En'yo}
\affiliation{Department of Physics, Kyoto University, 
Sakyo, Kyoto 606-8502, Japan}

\date{\today}

 \begin{abstract}
  We study the color correlation between
  static quark and antiquark ($q\bar q$) in the confined phase
  via reduced density matrices $\rho$ defined in color space.
  We adopt the standard Wilson gauge action and perform quenched
  calculations with the Coulomb gauge condition
  for reduced density matrices.
  The spatial volumes are $L^3 = 8^3$, $16^3$, $32^3$ and $48^3$, 
  with the gauge couplings $\beta = 5.7$, 5.8 and 6.0.
  Each element of the reduced density matrix in the sub space of quarks' color
  degrees of freedom of the $q\bar q$ pair is calculated
  from staples defined by link variables.
  As a result, we find that $\rho$ is well written
  by a linear combination of the strongly correlated
  $q\bar q$ pair state
  with the color-singlet component and the uncorrelated
  $q\bar q$ pair state
  with random color configurations. 
  We compute the Renyi entropies $S^{\rm Renyi}$
  from $\rho$ to investigate the $q\bar q$ distance dependence
  of the color correlation of the $q\bar q$ pair
  and find that the color correlation is quenched as the distance increases.
 \end{abstract}

\maketitle

\section{Introduction}
\label{Sec.Introduction}

Color confinement is one of the nonperturbative features of Quantum
ChromoDynamics (QCD), the fundamental theory of the strong interaction.
The static interquark potential ($q \bar q$ potential) in the confinement phase
exhibits a linearly rising potential in the large-separation limit
giving the diverging energy, and quarks cannot be isolated.
Such confining features have been studied and confirmed
in several approaches~\cite{Greensite:2011}.

The color confinement may be illustrated by the flux tube formation
between quark and antiquark.
A color flux tube which has a constant energy per length
is formed between (color singlet) $q \bar q$ pair and this tube gives the linearly rising
$q\bar q$ potential~\cite{Bali:1994de, Bornyakov:2004uv}.
Note that QCD is nonabelian gauge theory
and hence such gluon fluxes have colors.
In other words, the color charge first associated with a color-singlet
$q \bar q$ pair flows into interquark flux tube as the $q \bar q$ separation
is enlarged
keeping the total system color singlet
~\cite{Tiktopoulos:1976sj, Greensite:2001nx}.
If the color charge of the $q\bar q$ part 
and that of the gluon part are separately considered, 
this color transfer can be regarded as a color charge leak 
from $q\bar q$ part to the gluon part in association with the screening effect.
This color leak should depends on the $q\bar q$ distance and
would
be observed as the distance dependence of the color correlation between quark and antiquark.

Such color correlation of the $q\bar q$ pair
may be detected by entanglement entropy (EE)
defined by the reduced density matrix.
EE quantifies 
an entanglement between
degrees of freedom in purely quantum systems,
and have been utilized in variety of physical systems
~\cite{Itou:2015cyu,Aoki:2015bsa,Kanada-Enyo:2015ncq,Takayanagi:2012kg,Bennett:1995tk, Calabrese:2004eu, Vidal:2002rm, Amico:2007ag, Horodecki:2009zz, Bennett:1996gf, Wootters:1997id, Vidal:2002zz}.
If the $q\bar q$ pair's correlation is strong, the $q\bar q$ part is well decoupled from the gluon part
and there is no entanglement between the $q\bar q$ and gluon parts.
In other words, the color leak from $q\bar q$ part can be measured by EE.
In this paper, we define the reduced density matrix $\rho$ for a static $q\bar q$
pair in terms of color degrees of freedom.
The density matrix is reduced into
subspace of  $q\bar q$ color configurations
by integrating out the gluons' degrees of freedom,
which is simply done by averaging the density matrix components
over gauge configurations,
and compute entanglement entropy $S$ with the reduced density matrix.
Constructing a simple ansatz for the reduced density matrix $\rho$,
we investigate the dependence of $S$ on the interquark distance $R$.

In Sec.~\ref{Sec.Formalism}, we give the formalism to compute
the reduced density matrix $\rho$ of $q\bar q$ system and the entanglement entropy $S$
of it. The details of numerical calculations and ansatz for $\rho$
are also shown in Sec.~\ref{Sec.Formalism}.
Results are presented in Sec.~\ref{Sec.Results}.
Sec.~\ref{Sec.Summary} is devoted to the summary and concluding remarks.

\section{Formalism}
\label{Sec.Formalism}

\subsection{reduced 2-body density matrix and $q\bar q$ correlation}

The entanglement between two subsystems $A$ and $B$ can be
quantified with entanglement entropy (EE).
From the density matrix $\rho_{AB}$ for a whole system $A+B$,
the reduced density matrix $\rho_{A}$ is obtained as
$\rho_{A}={\rm Tr}_{B}\left(\rho_{AB}\right)$.
Here, ${\rm Tr}_B$ is taken over the degrees of freedom of the
subsystem $B$.
The entanglement entropy 
$S^{\rm EE}_{A}$
 of the subsystem $A$ is then defined as
 $S^{\rm EE}_{A}=-{\rm Tr}_{A}\left(\rho_{A}\log \rho_{A}\right)$
 in the functional form of the von Neumann entropy.
The density matrix $\rho_{A}$ defined for the reduced space (the subsystem A) can give a non-zero value of EE because 
a part of information is lost from the 
$\rho_{AB}$ for the full space by tracing out
degree of freedom (Dof) of the subsystem $B$. 
The EE is zero only 
in the case of  $\rho_A^2=\rho_A$ 
when the subsystems $A$ and $B$ are completely decoupled from each other (not entangled).

Since our interest is being focused on the static $q\bar q$ pair's
color correlations, we divide the whole {\it color-singlet} system
into ({\it possibly colored}) two subsystems,
static (anti)quarks ($Q$) and ``others''($G$),
and consider color DoF
of the subsystems ($Q=A$ and $G=B$).
Other DoF contains all the gluon's DoF
including the vacuum polarization by the sea quark's loop.

In the actual calculations,
we compute the reduced two-body density matrix $\rho$ in the subsystem $Q$
by taking into account static quark's color configuration only.
Thus defined density matrix is nothing but the reduced density matrix
$\rho_Q$ that is obtained integrating out the other DoF $G$
in the full density matrix $\rho_{QG}$;
$\rho_Q = {\rm Tr}_{G}\left(\rho_{QG}\right)$.

The reduced two-body density operator $\hat\rho(R)$ in a $q\bar q$ system
with the interquark distance $R$ is defined as
\begin{equation}
\hat\rho(R) = |\bar q(0) q(R)\rangle \langle \bar q(0) q(R)|.
\end{equation}
Here $|\bar q(0) q(R)\rangle$ represents a quantum state
in which the antiquark is located at the origin and the other quark lies at $x=R$.
The reduced density matrix components $\rho(R)_{ij,kl}$,
where $i$ ($j$) are quark's (antiquark's) color indices,
are expressed as
\begin{equation}
\rho(R)_{ij,kl} = \langle q_i(0)\bar q_j(R)|\hat\rho(R)|q_k(0)\bar q_l(R) \rangle.
\end{equation}
$\rho(R)$ is a $m\times m$ square matrix with the dimension $m=N_c^2$.
Note again that $\rho$ is defined using only quark's DoF and
gluon's wavefunction is not considered and 
then thus defined $\rho$ can be regarded as a reduced density matrix
where gluon's DoF are integrated out.

The  von Neumann entanglement entropy $S^{\rm VN}(R)$ 
for $q\bar q$ pair at a distance of $R$
can be computed with the reduced density matrix $\rho(R)$ as
\begin{equation}
S^{\rm VN} (R)
\equiv
-{\rm Tr} \ \rho(R) \log\rho(R)
=-\sum_{ij}\left[\rho(R)\log\rho(R)\right]_{ij},
\end{equation}
which can be regarded as an entanglement entropy
representing the correlation between static-quark pair (subsystem $Q$)
and other DoF (subsystem $G$).

In the actual computation of $S^{\rm VN}$,
one needs to diagonalize $\rho$ or approximate the logarithmic function.
In order to avoid such  numerically demanding processes,
we adopt Renyi entropy~\cite{Renyi:1970} for EE for detailed analysis.
Renyi entanglement entropy $S^{{\rm Renyi}-\alpha}$ of order $\alpha$ ($\alpha > 0$, $\alpha\neq 1$)
is given as
\begin{equation}
S^{{\rm Renyi}-\alpha} = \frac{1}{1-\alpha}\log {\rm Tr}\left(\rho^\alpha \right),
\end{equation}
with a reduced density matrix $\rho$.
Note that in the limit when $\alpha\rightarrow 1$,
it goes to von Neumann entropy as $S^{{\rm Renyi}-\alpha}\to S^{\rm VN}$. 
Renyi entanglement entropy is a kind of generalized entropies that quantify
uncertainty or randomness,
and used to measure entanglement in quantum information theory.
Since entanglement entropy is invariant under unitary transformations,
it enables representation independent analysis.
We use the second order Renyi entanglement entropy by taking $\alpha=2$, 
which is simply given by the squared $\rho(R)$ as  
\begin{equation}
S^{{\rm Renyi}-2} = -\log {\rm Tr}\left(\rho^2 \right).
\end{equation}


We here comment on the relationship between $q\bar q$ correlation
and the entanglement entropy.
Our main interest is the $q\bar q$ pair's color correlation defined
in the subsystem $Q$.
The whole pure state in $Q+G$ system can be written as
\begin{eqnarray}
\sum_\alpha |\alpha\rangle_Q \otimes |\alpha\rangle_G.
\end{eqnarray}
Here, $\alpha$ denotes all the possible color
states of the $q\bar q$ pair,
and total system is kept in a color singlet state.
When quark and antiquark's colors 
are strongly correlated forming a color singlet combination $|{\bm 1}\rangle_Q$
with
no color charge leak from $Q$ to $G$, 
the subsystems $Q$ and $G$ are well decoupled in the color space and therefore
the whole state can be expressed in a simple product of $Q$ and $G$ parts as
\begin{eqnarray}
\sum_{\alpha={\bm 1}} |\alpha\rangle_Q \otimes |\alpha\rangle_G
=
|{\bm 1}\rangle_Q \otimes |{\bm 1}\rangle_G.
\end{eqnarray}
In this strongly correlated case,
the entanglement entropy $S^{\rm EE}$ goes to zero,
since two subsystems $Q$ and $G$ decouple and 
the entanglement between subsystems $Q$ and $G$ vanishes.

On the other hand,
when $q\bar q$ pair's color charge leaks into inbetween gluons
and the color correlation between them decreases,
the whole state cannot be written in a separable form,
and $S$ would take a positive finite value as $S>0$.

\subsection{Ansatz for reduced density matrix $\rho_{ij,kl}(R)$}

Let us consider a possible functional form of the reduced density matrix 
$\rho_{ij,kl}(R)$ 
based on the simple ansatz that the contamination mixed 
to the correlated color singlet component 
is the random color component without any color correlation
between quark and antiquark of the $q\bar q$ pair.
We first define the density operator $\hat\rho_{{\bm s},{\bm s}}$
for quark and antiquark in a color singlet state
$|{\bm s}\rangle = \sum_i^{N_c} |\bar q_i q_i\rangle$
in the Coulomb gauge
as
\begin{equation}
\hat\rho_{{\bm s},{\bm s}} = |{\bm s}\rangle \langle {\bm s}|.
\end{equation}
In color SU($N_c$) QCD, 
the density operator $\hat\rho_{{\bm a}_i,{\bm a}_i}\ \ (i=1,2,..., N_c^2-1)$
for $q\bar q$ in an adjoint state
 $|{\bm a}_i\rangle\ \ (i=1,2,..., N_c^2-1)$
 is expressed as
\begin{equation}
\hat\rho_{{\bm a}_i,{\bm a}_i} = |{\bm a}_i\rangle \langle {\bm a}_i|
\ \ (i =1,2,...,N_c^2-1).
\end{equation}

In the limit $R\rightarrow 0$,
quark and antiquark are considered
to form a color-singlet state ($|{\bm s}\rangle$)
corresponding to the strong correlation limit,
and its density operator will be written as
\begin{equation}
 \hat \rho^{\rm 0} = \hat\rho_{{\bm s},{\bm s}}
 =
 {\rm diag}(1,0,...,0)_{\alpha{\rm -rep.}}
\end{equation}
Here, ``$\alpha{\rm -rep.}$'' means that the matrix is expressed
in terms of $q\bar q$'s color representation
with the vector set of $\{{\bm s},{\bm a_1},...{\bm a_8} \}$.
As $R$ increases,
it is expected that adjoint components mix into the singlet component
due to the QCD interaction.
We assume that
contamination mixed into the pure singlet (correlated) state 
is the uncorrelated state with random color configurations 
where $N_c^2$ components mix with equal weights. 
The density operator for such the random state is given as 
\begin{eqnarray}
 \hat \rho^{\rm rand}
&=&
 \frac{1}{N_c^2}\hat\rho_{{\bm s},{\bm s}}
 +
 \frac{1}{N_c^2}\hat\rho_{{\bm a}_1,{\bm a}_1}
 +
 \frac{1}{N_c^2}\hat\rho_{{\bm a}_2,{\bm a}_2}
 +...
\nonumber \\
&=&
\frac{1}{N_c^2}{\hat I}
 =
\frac{1}{N_c^2}{\rm diag}(1,1,...,1)_{\alpha{\rm -rep.}}
\end{eqnarray}

Letting the fraction of the original (maximally correlated)
singlet state being $F(R)$
and that of the mixing (random) components being $(1-F(R))$,
the density operator in this ansatz is written as
\begin{eqnarray}
 \hat\rho_{\rm ansatz}(R)
  &=&
  F(R)\hat\rho^{\rm 0}+(1-F(R))\hat\rho^{\rm rand}.
\end{eqnarray}
The matrix elements of $\hat\rho_{\rm ansatz}(R)$ in the $\alpha$-representation
are explicitly written as
\begin{widetext}
\begin{eqnarray}
 \hat\rho_{\rm ansatz}(R)
  &=&
  F(R)\hat\rho^{\rm 0}+(1-F(R))\hat\rho^{\rm rand}
  \\
 &=&
  {\rm diag}\left(
	     F(R)+\frac{1}{N_c^2}(1-F(R)),\frac{1}{N_c^2}(1-F(R))
	     ,...,
	     \frac{1}{N_c^2}(1-F(R))
	    \right)_{\alpha{\rm -rep.}}
\\
&=&
\begin{pmatrix}
F(R)+\frac{1}{N_c^2}(1-F(R)) & 0 &  & \cdots &  0
\\
0  & \frac{1}{N_c^2}(1-F(R)) &  &  &  \vdots
\\
\vdots  & & \ddots &  & \vdots
\\
\vdots   &  &  &  &  0
\\
0  & \cdots &  &  0 &  \frac{1}{N_c^2}(1-F(R))
\\
\end{pmatrix}
_{\alpha-{\rm rep.}}
\end{eqnarray}

\end{widetext}
When $N_c=3$, 
\begin{eqnarray}
 \begin{cases}
 \rho(R)_{{\bf 8}_1,{\bf 8}_1}
 =
 \rho(R)_{{\bf 8}_2,{\bf 8}_2}
 =...=
 \rho(R)_{{\bf 8}_8,{\bf 8}_8}
\equiv
 \rho(R)_{{\bf 8},{\bf 8}}
\\  
 \rho(R)_{\alpha,\beta}=0\ \ ({\rm for}\ \alpha\neq\beta)
 \end{cases}
\label{Eq.conditions}
\end{eqnarray}
would be satisfied at any $R$ in this ansatz.
The first relation should be satisfied due to the color SU(3) symmetry.
The second, which means that the off-diagonal components are all zero,
comes from the ansatz of the random state.
The normalization condition ${\rm Tr}\rho = 1$
is trivially satisfied in this ansatz as
\begin{eqnarray}
 \rho(R)_{{\bf 1},{\bf 1}}
+
(N_c^2-1)
 \rho(R)_{{\bf 8},{\bf 8}}
=1.
\end{eqnarray}

In the strong correlation limit
when $q\bar q$ pair's color forms $|{\bm 1}\rangle$,
$F(R)=1$.
On the other hand, in the random limit
when quarks' colors are screened, $F(R)=0$.

\subsection{Lattice QCD formalism}

Let the site on the lattice
${\bm r}=(x,y,z,t)=x{\bm e_x}+y{\bm e_y}+z{\bm e_z}+t{\bm e_t}$
and $\mu$-direction ($\mu=x,y,z,t$) link variables
being $U_\mu({\bm r})$.
With a lower staple $S^L(R,T)$ representing
$q\bar q$ pair creation and propagation and 
an upper staple $S^U(R,T)$ for $q\bar q$ pair annihiration
that are defined as
\begin{eqnarray}
 S^L_{ij}(R,T)\equiv
\left(
\prod_{t=-1}^{-T} U_{t}^\dagger(t{\bm e_t})
\prod_{x=0}^{R-1}  U_{x}(x{\bm e_x}-T{\bm e_t})\right. \nonumber
\\ 
\left. \times \prod_{t=-T}^{-1} U_{t}(R{\bm e_x}+t{\bm e_t})
\right)_{ij},
\end{eqnarray}
\begin{eqnarray}
 S^U_{ij}(R,T)\equiv
\left(
\prod_{t=0}^{T-1} U_{t}(t{\bm e_t})
\prod_{x=0}^{R-1}  U_{x}(x{\bm e_x}+T{\bm e_t})\right. \nonumber
\\
\left. \times
\prod_{t=T-1}^{0} U_{t}^\dagger(R{\bm e_x}+t{\bm e_t})
\right)_{ij},
\end{eqnarray}
we define $L_{ij}(R,T)$ as
\begin{equation}
 L_{ij,kl}(R,T)\equiv
  S^U_{ij}(R,T)S^{L\dagger}_{kl}(R,T).
\end{equation}
When the euclidean time separation $T$ is large enough
and excited state contributions can be ignored,
$\langle L_{ij,kl}(R,T) \rangle$ is expressed as
\begin{eqnarray}
&&
\langle L_{ij,kl}(R,T) \rangle
\nonumber \\
&=&
C
\langle q(0)\bar q(R)|
e^{- \hat{H} T} |q_i(0) \bar q_j(R)\rangle
\nonumber \\
&\times&
\langle \bar q_k(0) q_l(R)|e^{- \hat{H} T}
| q(0)\bar q(R)\rangle
\nonumber \\
&=&
Ce^{- 2E_0 T}\langle q(0)\bar q(R)|q_i(0) \bar q_j(R)\rangle\langle \bar
q_k(0) q_l(R)| q(0)\bar q(R)\rangle
\nonumber \\
&=&
Ce^{- 2E_0 T}\rho(R)_{ij,kl},
\end{eqnarray}
where $E_0$ is the ground-state energy.
Normalizing $\langle L(R,T) \rangle$ so that
${\rm Tr}\ \langle L(R,T) \rangle =\sum_{ij} \langle L_{ij,ij}(R,T)
\rangle = 1$,
we obtain $\rho(R)$ whose trace is unity (${\rm Tr}\ \rho(R)=1$).

Once we obtain $\rho(R)$, Renyi entropy of order $\alpha$
as a function of $R$ is obtained as
\begin{equation}
S^{{\rm Renyi}-\alpha}(R) = \frac{1}{1- \alpha}\log {\rm Tr}\left(\rho(R)^\alpha \right).
\end{equation}

\subsection{Lattice QCD parameters}

We adopt the standard Wilson gauge action and perform quenched
calculations for reduced density matrices of static quark and antiquark ($q\bar q$) systems.
The gauge configurations are generated
on the spatial volumes $L^3 = 8^3$, $16^3$, $32^3$ and $48^3$, 
with the gauge couplings $\beta = 5.7$, 5.8 and 6.0.
All the gauge configurations are gauge-fixed with the Coulomb gauge condition.
The parameters adopted in the present work are summarized in Table.~\ref{params}.

\begin{table}[h]
\begin{tabular}{c c c c}
\hline
$\beta$ & $a$ [fm] & $L^3$ & $L^3$ [fm$^3$]\\
\hline
5.7 & 0.18 & $8 ^3$ & 1.44$^3$ \\
5.7 & 0.18 & $16^3$ & 2.88$^3$ \\
5.7 & 0.18 & $32^3$ & 5.75$^3$ \\
5.7 & 0.18 & $48^3$ & 8.64$^3$ \\
5.8 & 0.14 & $16^3$ & 2.24$^3$ \\
5.8 & 0.14 & $32^3$ & 4.48$^3$ \\
5.8 & 0.14 & $48^3$ & 6.72$^3$ \\
6.0 & 0.10 & $16^3$ & 1.60$^3$ \\
6.0 & 0.10 & $32^3$ & 3.20$^3$ \\
6.0 & 0.10 & $48^3$ & 4.80$^3$ \\
\hline
\end{tabular}
\caption{\label{params}
 Lattice QCD parameters. Coupling $\beta$, lattice spacing $a$, spatial volume $L^3$
 in the lattice unit and the physical unit.
}
\end{table}

\section{Lattice QCD results}
\label{Sec.Results}

\subsection{Ground-state dominance}

\begin{figure}[h]
\includegraphics[width=08cm]{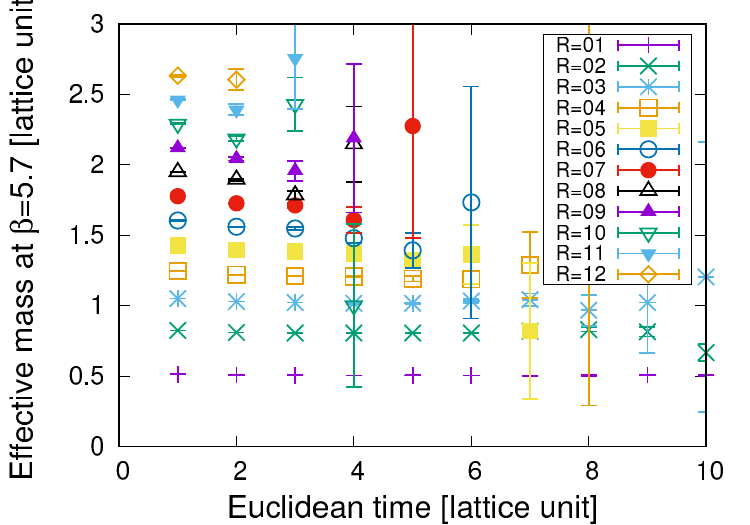}
\caption{\label{efmass57}Effective energy plot as a function of the Euclidean time
 separation at $\beta=5.7$. $R$ denotes the interquark distance in
 lattice unit.}
\end{figure}

\begin{figure}[h]
\includegraphics[width=08cm]{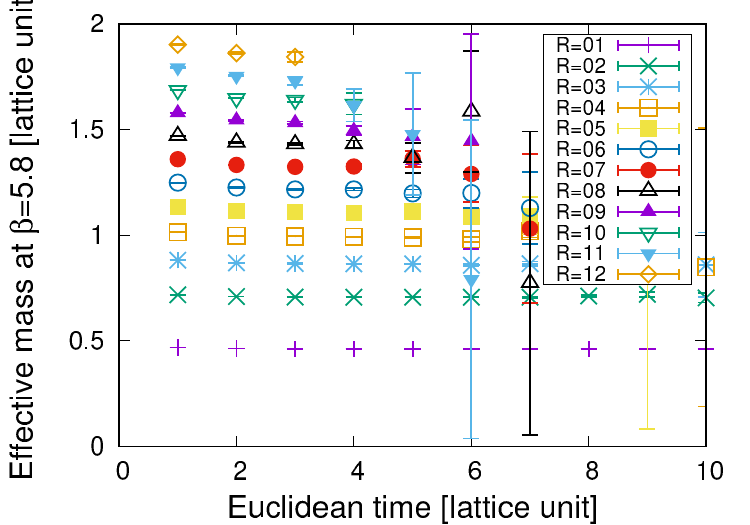}
\caption{\label{efmass58}Effective energy plot as a function of the Euclidean time
 separation at $\beta=5.8$. $R$ denotes the interquark distance in
 lattice unit.}
\end{figure}

\begin{figure}[h]
\includegraphics[width=08cm]{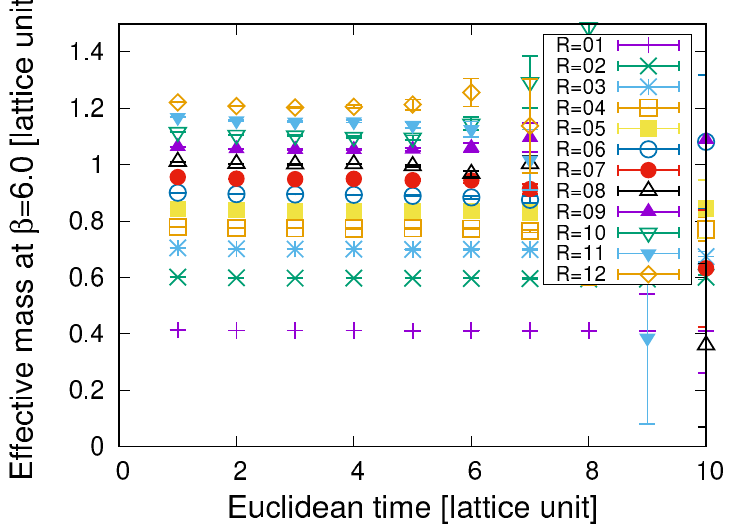}
\caption{\label{efmass60}Effective energy plot as a function of the Euclidean time
 separation at $\beta=6.0$. $R$ denotes the interquark distance in
 lattice unit.}
\end{figure}

In order to confirm the ground-state dominance,
we investigate the static quark and antiquark potential.
In Figs.~\ref{efmass57}, \ref{efmass58} and \ref{efmass60}, 
we show the effective energy plots for static $q\bar q$ systems with several
interquark distances $R$ as a function of the Euclidean time separation $T$.
For all the interquark distances $R$,
effective energies show plateaux against $T$ at $T\geq 2$
and it is confirmed that ground-state saturation is ensured at $T\geq 2$.
Hereafter, we adopt normalized reduced density matrix $\rho(R,2)$ measured with $T=2$
for $\rho_{ij,kl}(R)$;
$\rho_{ij,kl}(R)\equiv \rho_{ij,kl}(R,2)/{\rm Tr}\ \rho(R,2)$.


\subsection{Reduced density matrix elements}

\begin{figure}[h]
\begin{center}
\includegraphics[width=08cm]{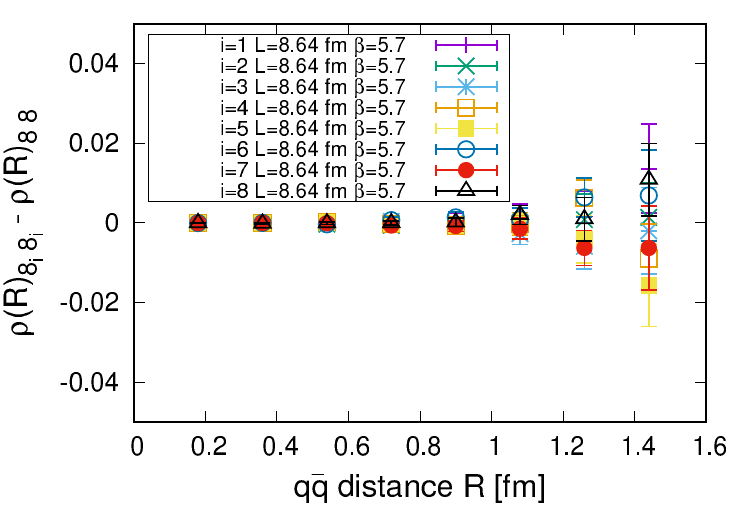}
\end{center}
 \caption{\label{D(R)}The deviation of each component $\rho(R)_{{\bm
 8}_i,{\bm 8}_i}$ from the averaged value $\rho(R)_{{\bf 8}, {\bf 8}}=\frac{1}{N_c^2-1}\sum_i \rho(R)_{{\bf
 8}_i,{\bf 8}_i}$ is plotted as a
 function of the interquark distance. They are evaluated
 at $\beta=5.7$ and $L=48$. All the values are consistent with zero
 within the errors.
 }
\end{figure}

In this subsection, 
we take a detailed look at the reduced density matrix elements obtained with lattice QCD.
In order to see the validity of the first condition in Eq.(\ref{Eq.conditions}),
we define the average
\begin{equation}
\rho(R)_{{\bf 8}, {\bf 8}}=\frac{1}{N_c^2-1}\sum_i \rho(R)_{{\bf
 8}_i,{\bf 8}_i}
\end{equation}
and the deviation
\begin{equation}
 D_i(R) = (\rho(R)_{{\bf 8}_i,{\bf 8}_i}-\rho(R)_{{\bf 8},{\bf 8}}).
\end{equation}
In Fig.~\ref{D(R)}, $D_i(R)$ ($1\leq i\leq 8$) are plotted as a function of the interquark
distance. All the values are consistent with zero and it is confirmed
that the first condition is satisfied for all the $R$ and $i$ within statistical errors.
Hereafter, the octet components of $\rho(R)$ is represented
by the averaged value $\rho(R)_{{\bf 8}, {\bf 8}}$.

The second condition in Eq.(\ref{Eq.conditions})
is the assumption in the ansatz.
To see to what extent this assumption is valid in the actual reduced density
matrices, we define following two independent components.
\begin{equation}
 \rho(R)_{{\bf 1},{\bf 8}_1}
 =
 -\frac{1}{\sqrt{3}}
 \left(
\rho(R)_{11,12}+\rho(R)_{22,12}+\rho(R)_{33,12}
 \right),
\end{equation}
\begin{equation}
 \rho(R)_{{\bf 8}_3,{\bf 8}_4}
 =
\rho(R)_{21,13}.
\end{equation}
\begin{figure}[h]
\begin{center}
\includegraphics[width=08cm]{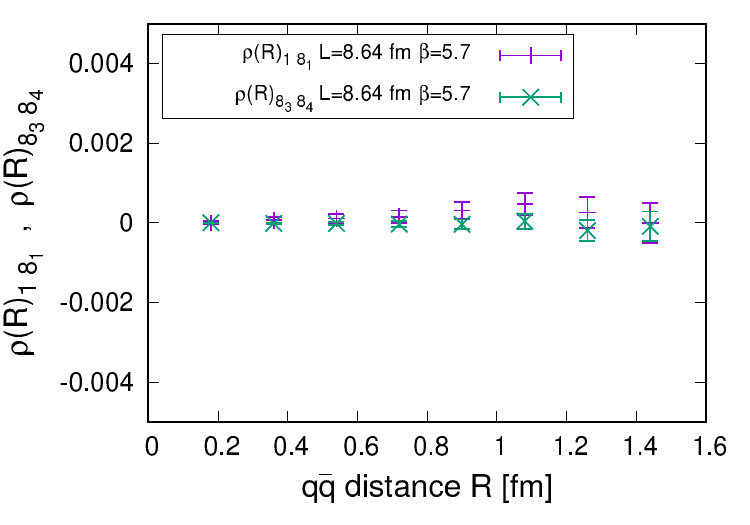}
\end{center}
 \caption{\label{CK2}
 In order to see the magnitudes of the off-diagonal components,
 two independent off-diagonal components
$\rho(R)_{{\bf 1},{\bf 8}_1}$ and $\rho(R)_{{\bf 8}_3,{\bf 8}_4}$ are
 plotted as a function of the interquark distance.
 They are consistent with zero and we conjecture that
 the off-diagonal components of the reduced density matrix $\rho(R)$
 are consideably small.} 
\end{figure}
$\rho(R)_{{\bf 1},{\bf 8}_1}$ and $\rho(R)_{{\bf 8}_3,{\bf 8}_4}$ are plotted
in Fig.~\ref{CK2}.
We find that they are consistent with zero and we conjecture that
the off-diagonal components of the reduced density matrix $\rho(R)$ are
considerably small.
From these analyses, we can conclude that
the reduced density matrix $\rho(R)$ obtained with lattice calculations
in the static $q\bar q$ system
is expressed by the ansatz with high accuracy.
Indeed, even when we replace the octet components
and the off-diagonal components
of $\rho(R)$ with the average $\rho(R)_{{\bf 8}, {\bf 8}}$
and with zero by hand,
all the results remain almost unchanged.

\subsection{$R$ dependence of $F(R)$}

Taking into account the normalization condition
\begin{equation}
 \rho(R)_{{\bf 1},{\bf 1}}
+
(N_c^2-1)
 \rho(R)_{{\bf 8},{\bf 8}}
=1,
\end{equation}
the independent quantity at a given R is only  $\rho_{{\bf 8}, {\bf 8}}$,
and we can compute the fraction $F(R)$ of the remaining correlated $q\bar q$
component as,
\begin{equation}
 F(R) 
 =\rho(R)_{{\bf 1},{\bf 1}}-\rho(R)_{{\bf 8},{\bf 8}}
= 1-N_c^2 \rho(R)_{{\bf 8},{\bf 8}}.
\end{equation}
When the $q\bar q$ system forms a random state with 
no color correlation between $q$ and $\bar q$,
the calculated $\rho(R)$ equals to $\hat\rho^{\rm rand}$ and gives $F(R)=0$.
\begin{figure}[h]
\begin{center}
\includegraphics[width=08cm]{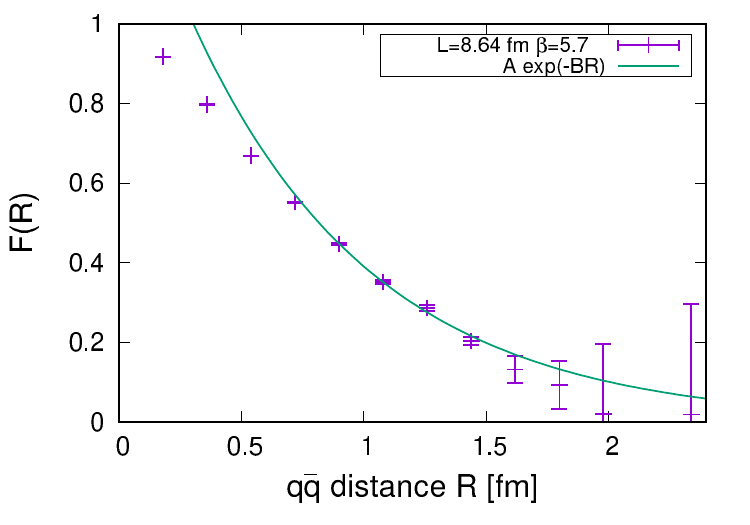}
\includegraphics[width=08cm]{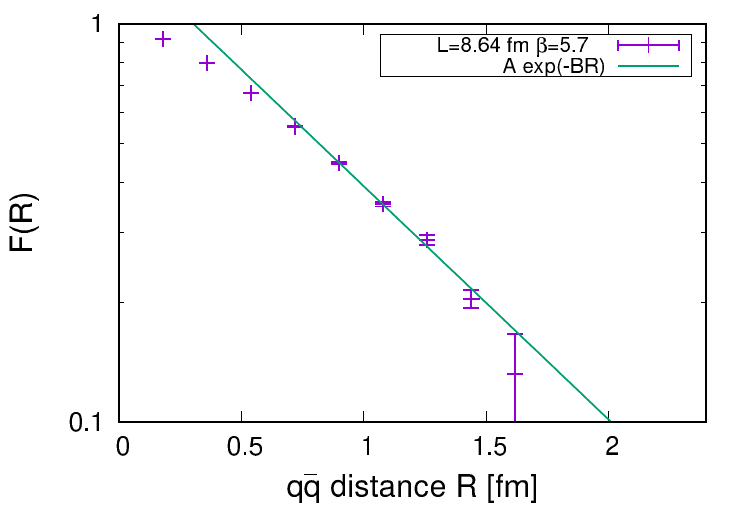}
\end{center}
 \caption{
 \label{F(R)} $F(R)$ is plotted as a function of the interquark
 distance $R$ in the upper panel. $F(R)$ monotonously decreases
 and approaches zero. The lower panel shows the log plot for $F(R)$.
 The fit function,
 $F(R)  = A \exp(-B R)$ with $A=1.505$ and $B=1.347$\ ${\rm fm}^{-1}$, 
 is shown as a solid line.
 }
\end{figure}
In the upper panel in Fig.~\ref{F(R)},
$F(R)$ is plotted as a function of the interquark
distance $R$. 
$F(R)$ linearly decreases at small $R$,
and exponentially approaches zero at large $R$,
which can be also seen in the lower panel (logarithmic plot of $F(R)$).

The exponential decay of the $q\bar q$ correlation
indicates
the color screening effects due to inbetween gluons.
We fit $F(R)$ with an exponential function as
\begin{equation}
F(R)  = A \exp(-B R)
\end{equation}
and extract the ``screening mass'' $B$.
In Fig.~\ref{params},
the fitted parameters $A$ and $B$ are plotted
as functions of the spatial lattice size $L$.
The plot includes all the data obtained at $\beta$=5.7, 5.8 and 6.0
so that one can see the $\beta$ (lattice spacing) dependence.
While a tiny deviation is found among three $\beta$'s,
all the data seem lie in a monotoneous line,
which means the systematic errors for $A$ and $B$ mainly
arise from the lattice size $L$.
For $L>5$ fm, the fitted values are stable,
and $A$ and $B$ are deteremined as
\begin{eqnarray}
A&=&1.505(49) \\
B&=&1.347(35) \ {\rm fm}^{-1} = 265(7) \ {\rm MeV}
\end{eqnarray}
from $F(R)$ obtained in the largest volume.

\begin{figure}[h]
\includegraphics[width=08cm]{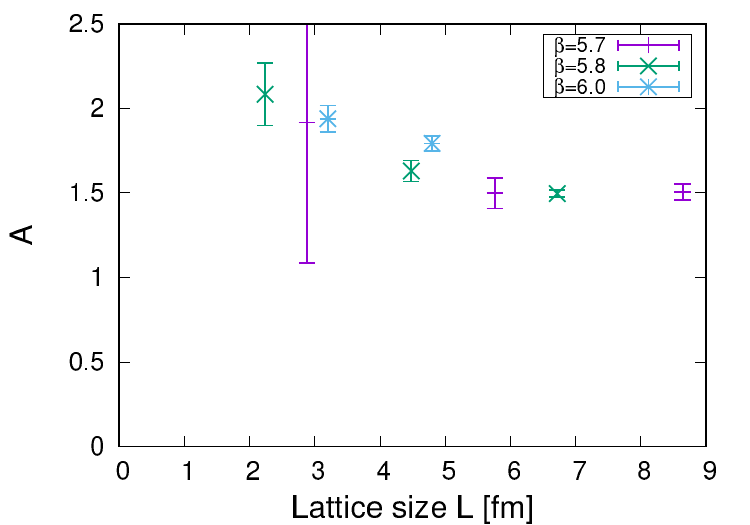}
\includegraphics[width=08cm]{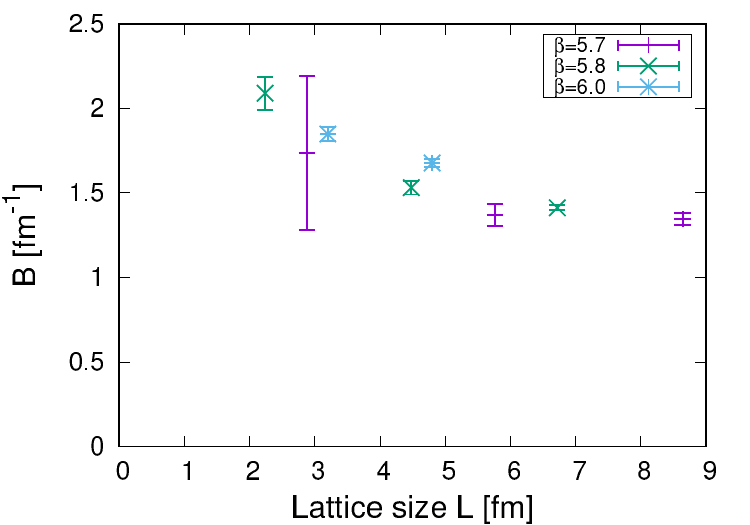}
 \caption{\label{params}
 The fitted parameters $A$ and $B$ plotted
 as functions of the spatial lattice size $L$.
 }
\end{figure}

\begin{figure}[h]
\begin{center}
\includegraphics[width=08cm]{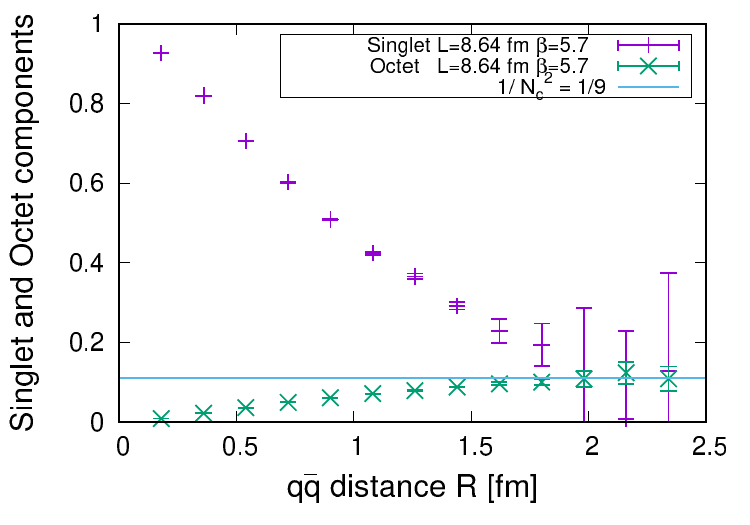}
\end{center}
 \caption{\label{18components}
 The singlet component $\rho(R)_{{\bf 1},{\bf 1}}$ and the averaged
 octet component $\rho(R)_{{\bf 8},{\bf 8}}$ are plotted as a function
 of the interquark distance $R$.
 Both are approaching $1/Nc^2=1/9$ at large $R$.
 }
\end{figure}

In Fig.~\ref{18components},
the singlet component $\rho(R)_{{\bf 1},{\bf 1}}$ and the averaged
octet component $\rho(R)_{{\bf 8},{\bf 8}}$ are plotted as a function
of the interquark distance $R$.
One finds that both components approach 
$\rho(R)_{{\bf 1},{\bf 1}}=\rho(R)_{{\bf 8},{\bf 8}}=\frac{1}{N_c^2}=\frac19$
at large $R$,
which ensures that the reduced density matrix at large interquark separation $R$
is governed by the random component
$\hat\rho^{\rm rand}$ and the original correlated state $\hat\rho^0$ vanishes.

\subsection{Finite volume effects}

Within the present numerical accuracy,
the only independent quantity in the reduced density matrix $\rho(R)$
is $\rho_{{\bf 8},{\bf 8}}$,
and all the finite volume effects are reflected in
$F(R)= 1-N_c^2 \rho(R)_{{\bf 8},{\bf 8}}$.
\begin{figure}[h]
\begin{center}
\includegraphics[width=08cm]{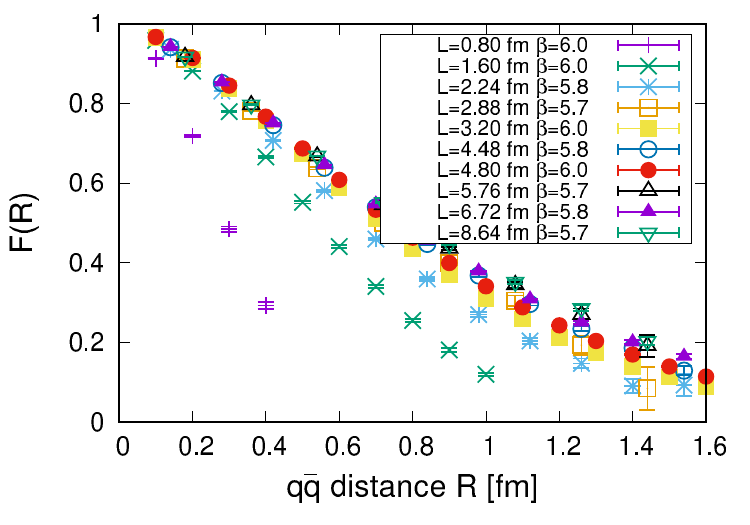}
\end{center}
 \caption{\label{F(R)voldep}
 $F(R)$ for different $L$ (lattice size) are plotted
 as a function of the interquark distance $R$.
 At $L>5.0 {\rm fm}$, $F(R)$ shows almost no volume dependence
 and $\rho(R)$ is safe from the finite volume effects at this $L$ range.} 
\end{figure}

In Fig.~\ref{F(R)voldep},
$F(R)$ for several $L$ (lattice size)
and $\beta$ (lattice spacing) are plotted
as a function of the interquark distance $R$.
At $L>5.0$ fm, $F(R)$ shows almost no volume dependence
and $\rho(R)$ is safe from the finite volume effects at this $L$ range.
When the lattice size $L$ is small,
$F(R)$ rapidly decreases with increasing $R$.
On the other hand, $\beta$ dependence seems smaller than
the finite volume effect.
The systematic errors mainly comes from the finite size effect.

This finite volume effect
would be due to the periodic boundary condition,
with which identical $q\bar q$-systems exist
with the period $L$.
Quark and antiquark ($q(0)\bar q(R)$) separated by $R$ in a system can
also form color singlet pairs
with quarks that are separated with the distance $L-R$,
which additionally enters in $\rho(R)$
as a random mixture decreasing $F(R)$.

\subsection{Entanglement entropy}

In the following, we consider $\alpha=2$ case for the evaluation
of the EE.
(We will go back to $S^{\rm VN}$ in the latter part of this section.)
The $S^{\rm Renyi-2}(R)$ is correctly calculated from 
the trace of the squared reduced density matrix $\rho(R)$
as
\begin{equation}
S^{{\rm Renyi}-2} =-\log {\rm Tr}(\rho(R)^2).
\end{equation}
Taking into account that
${\rm Tr}(\rho(R))=1$, 
the maximum of $S^{{\rm Renyi}-2}$ 
is obtained when
all the $N_c^2$ diagonal elements are equal to $1/N_c^2$
in the diagonal representation of $\rho(R)$.
From the representation invariance of $S$,
the maximum value of $S$ is proved to be
\begin{equation}
{\rm max}\left[S^{\rm Renyi-2}(R)\right]
=2\log N_c.
\end{equation}

\begin{figure}[h]
\begin{center}
\includegraphics[width=08cm]{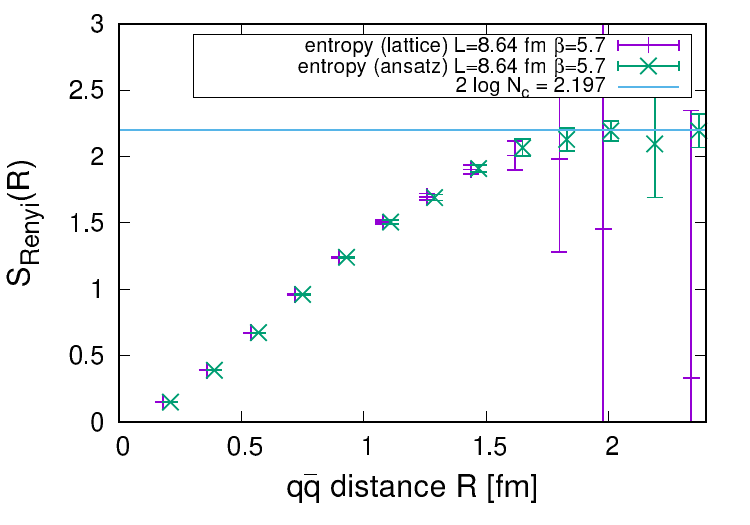}
\end{center}
 \caption{\label{SRenyi2}
$S^{\rm Renyi-2}_{\rm lattice}(R)$ obtained
 from the original reduced density matrix $\rho(R)$ and
 $S^{\rm Renyi-2}_{\rm ansatz}(R)$ obtained using the ansatz
 are plotted as a function of the interquark distance $R$.}
\end{figure}

In Fig.~\ref{SRenyi2},
$S^{\rm Renyi-2}(R)$ calculated
with the $\rho(R)$ obtained on the lattice
are plotted as $S^{\rm Renyi-2}_{\rm lattice}(R)$.
$S^{\rm Renyi-2}_{\rm lattice}(R)$ approaches $2\log N_c$
as $R$ increases,
which indicates that
$\rho(R)$ is described by the random component $\hat \rho^{\rm rand}$
in the large $R$ limit.

In the ansatz, the density matrix $\rho_{\rm ansatz}(R)$ is a diagonal matrix 
and ${\rm Tr}(\rho_{\rm ansatz}(R)^2)$ is given by $F(R)$ as
\begin{equation}
{\rm Tr}(\rho_{\rm ansatz}(R)^2)=F(R)^2+\frac{1}{N_c^2}-\frac{F(R)^2}{N_c^2}.
\end{equation}
Then $S^{\rm Renyi-2}_{\rm ansatz}(R)$,
the Renyi entropy evaluated using the ansatz,
 is expressed as
\begin{eqnarray}
S^{\rm Renyi-2}_{\rm ansatz}(R)
&=& -\log {\rm Tr}(\rho_{\rm ansatz}(R)^2)
\nonumber \\
&=& -\log\left(F(R)^2+\frac{1}{N_c^2}-\frac{F(R)^2}{N_c^2}\right).
\end{eqnarray}

Fig.~\ref{SRenyi2} shows
 $S^{\rm Renyi-2}_{\rm ansatz}(R)$ obtained using the ansatz
plotted as a function of the interquark distance $R$.
$S^{\rm Renyi-2}_{\rm ansatz}(R)$ approaches $2\log N_c$ at large $R$,
which again confirms that
$F(R)$ goes to zero and 
$\rho_{\rm ansatz}(R)$ is expressed by the random elements $\hat\rho^{\rm rand}$
in the $R\rightarrow \infty$ limit.
The remarkable fact is that
$S^{\rm Renyi-2}_{\rm lattice}(R)$ and $S^{\rm Renyi-2}_{\rm ansatz}(R)$
are almost identical for all $R$,
which indicates that the reduced density matrix $\rho(R)$ can be very well expressed
by the ansatz.
\begin{figure}[h]
\begin{center}
\includegraphics[width=08cm]{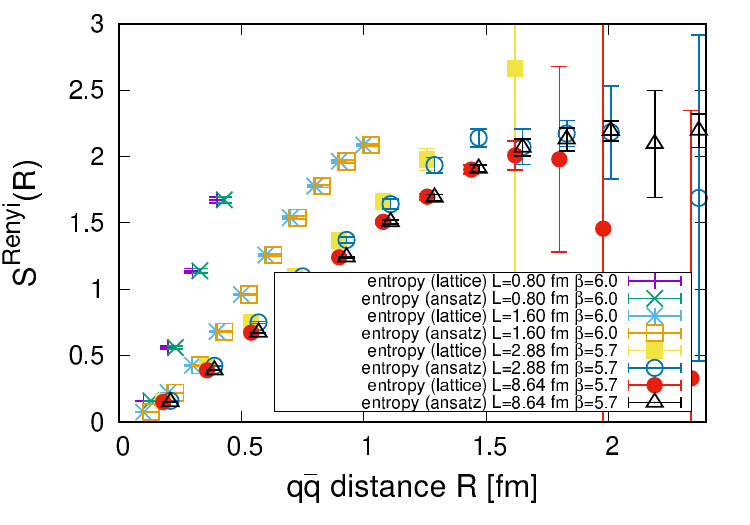}
\end{center}
 \caption{\label{SRenyi2voldep}
 $S_{\rm ansatz}^{\rm Renyi-2}$ and $S_{\rm lattice}^{\rm Renyi-2}$,
 which are obtained from the original reduced density matrix $\rho(R)$ and that
 obtained using the ansatz,
 are plotted as a function of $R$.
 }
\end{figure}

$S_{\rm ansatz}^{\rm Renyi-2}$ and $S_{\rm lattice}^{\rm Renyi-2}$
for different lattice sizes $L$
are plotted as a function of $R$ in Fig.~\ref{SRenyi2voldep}.
As expected, when $L<5.0$ fm, the finite volume effects are rather large,
and $S_{\rm ansatz}^{\rm Renyi-2}$ and $S_{\rm lattice}^{\rm Renyi-2}$ are
both affected.
On the other hand, for all the $L$, 
$S_{\rm ansatz}^{\rm Renyi-2}\simeq S_{\rm lattice}^{\rm Renyi-2}$ is found
and the ansatz is valid with a good accuracy
even when the finite volume effects are large.

It is well known that any averaging leads to the growth of the entropy.
The reduced density-matrix components are averaged
in the ansatz and one may think 
$S_{\rm ansatz}^{\rm Renyi-2} > S_{\rm lattice}^{\rm Renyi-2}$
should be observed.
Although such tendency can be sometimes seen in figures,
statistical errors are much larger and 
both data are consistent with each other within the present statistics.

\begin{figure}[h]
\begin{center}
\includegraphics[width=08cm]{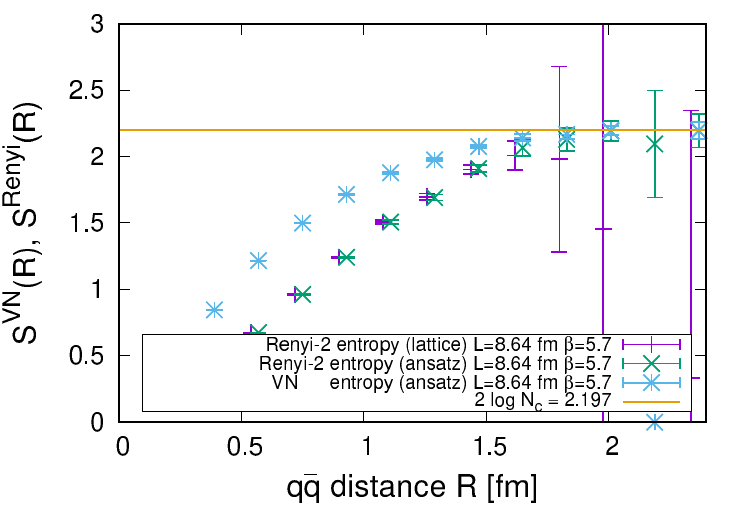}
\end{center}
 \caption{\label{SVN}
$S_{\rm ansatz}^{\rm VN}$, $S_{\rm ansatz}^{\rm Renyi-2}$ and $S_{\rm lattice}^{\rm Renyi-2}$,
 which are obtained from the original reduced density matrix $\rho(R)$ and that
 obtained using the ansatz, are plotted as a function of $R$.
 }
\end{figure}

Finally, we show the von Neumann entropy $S^{\rm VN}$ based on the ansatz.
The direct calculation of $S^{\rm VN}$ from the reduced density matrix on the
lattice is numerically demanding.
Instead of such a straightforward approach,
we take an alternative way to calculate $S^{\rm VN}$
with an approximation using $\rho_{\rm ansatz}$
from the ansatz
as
\begin{equation}
S^{\rm VN}_{\rm ansatz} = -{\rm Tr}\left(\rho_{\rm ansatz}\log\rho_{\rm ansatz}\right).
\end{equation}
$\rho$ evaluated on the lattice coincides with
$\rho_{\rm ansatz}$ with high accuracy as shown above,
and $S^{\rm VN}_{\rm ansatz}$ is expected to be a good approximation
of $S^{\rm VN}$.
Now the reduced density matrix in the $\alpha$-representation has been found to be diagonal
and then $S^{\rm VN}(R)$ is easily computed as
\begin{widetext}
\begin{eqnarray}
S^{\rm VN}_{\rm ansatz} (R)
=
&-&
\left(F(R)+\frac{1}{N_c^2}(1-F(R))\right)
\log \left(F(R)+\frac{1}{N_c^2}(1-F(R))\right) \nonumber
\\
&-&
(N_c^2-1) \left(\frac{1}{N_c^2}(1-F(R))\right)
\log \left(\frac{1}{N_c^2}(1-F(R))\right).
\end{eqnarray}
\end{widetext}
Fig.~\ref{SVN} shows $S^{\rm VN}_{\rm ansatz} (R)$ as a function of $R$,
and $S^{\rm Renyi-2}_{\rm lattice} (R)$ and $S^{\rm Renyi-2}_{\rm ansatz} (R)$
are also plotted for reference.
$S^{\rm VN}_{\rm ansatz} (R)$ increases towards $2\log N_c$
faster than $S^{\rm Renyi-2} (R)$
as the VN EE is a more sensitive measure of the entanglement
than the Renyi-2 EE in general.
As $R$ increases, the reduced density matrix $\hat\rho$ is dominated by the random contribution $\hat\rho^{\rm rand}$, and all the matrix elements are equipartitioned in this limit giving the maximum value of entropy.

\begin{figure}[h]
\begin{center}
\includegraphics[width=08cm]{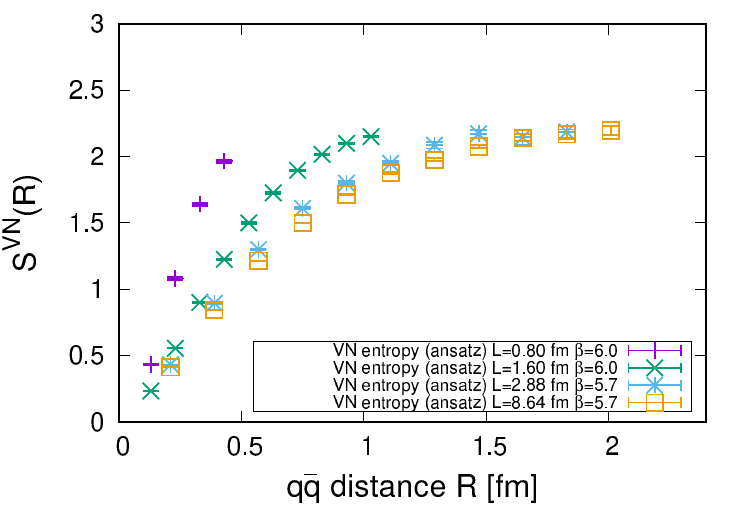}
\end{center}
 \caption{\label{SVNvoldep}
$S_{\rm ansatz}^{\rm VN}$ obtained using the ansatz are plotted as a
 function of $R$
 for different lattice size $L$.
 }
\end{figure}

In order to see the finite volume effects,
we plot 
$S_{\rm ansatz}^{\rm VN}$ as a function of $R$ in Fig.~\ref{SVNvoldep}.
The tendency that $S$ is increased
by the finite size effects remains unchanged.

\section{Summary and concluding remarks}
\label{Sec.Summary}

We have studied the color correlation
of static quark and antiquark ($q\bar q$) systems in the confined phase
from the viewpoint of the entanglement entropy (EE)
defined by reduced density matrices $\rho$ in color space.
We have adopted the standard Wilson gauge action and performed quenched
calculations for density matrices.
The gauge couplings are $\beta = 5.7$, 5.8 and 6.0, and
the spatial volumes are $L^3 = 8^3$, $16^3$, $32^3$ and $48^3$.
In order to evaluate each component of $\rho_{ij,kl}$,
all the gauge configurations are Coulomb-gauge fixed.
We have also proposed an ansatz for the reduced density matrix $\rho$,
in which $\rho$ is written
by a sum of
the color-singlet (correlated) state $|{\bm 1}\rangle\langle{\bm 1}|$
and random (uncorrlated)
elements $|{\bm 1}\rangle\langle{\bm 1}|$,
$|{\bm 8_i}\rangle\langle{\bm 8_i}|$ ($i=1,..,N_c^2-1$)
induced by the QCD interaction.

We have quantitatively evaluated the $q\bar q$ correlation
by means of the entanglement entropy constructed
from the reduced density matrix $\rho$.
We have adopted the von Neumann entropy $S^{\rm VN}$
and the Renyi entropy of the order $\alpha$ $S^{{\rm Renyi}-\alpha}$
for the evaluation of EE.
Especially when $\alpha$ is an integer,
$S^{{\rm Renyi}-\alpha}$
can be computed easily from the density matrix product $\rho^\alpha$,
and we need no diagonalization of $\rho$.
Note that color indices in EEs are all contracted,
and color-correlation measurement by means of EEs
can be performed in a gauge (representation) independent way.

As a result, we have found that
the reduced density matrix $\rho$
can be reproduced well with the ansatz:
The reduced density matrix $\rho$
consists of the color-singlet (correlated) state
$|{\bm 1}\rangle\langle{\bm 1}|$
when $q\bar q$ distance is small,
and random (uncorrlated)
diagonal elements $|{\bm 1}\rangle\langle{\bm 1}|$,
$|{\bm 8_i}\rangle\langle{\bm 8_i}|$ ($i=1,..,N_c^2-1$)
are equally mixed as $q\bar q$ distance is increased.
The $q\bar q$ color correlations
have been found to be well quantified by entanglement entropies,
and we conclude that
entanglement entropy can be a gauge independent measure for color correlations.

\appendix
\section{$\alpha$-representation and $ij$-representation}

In SU(3) QCD, $q\bar q$ state in $\alpha$-representation,
$|{\bf 1} \rangle$ and $|{\bf 8}_i \rangle$ ($i=1,2,..,N_c^2-1$),
can be expressed by states in $ij$-representation,
$|\bar q_i q_j \rangle$, as following.

\begin{eqnarray*}
&&
|{\bf 1} \rangle
=
\frac{1}{\sqrt{3}}
\left(
\sum_i
|\bar q_i q_i \rangle
\right)
,
\\
&&
|{\bf 8}_1 \rangle
=
-|\bar q_1 q_2 \rangle
,
\\
&&
|{\bf 8}_2 \rangle
=
-\frac{1}{\sqrt{2}}
\left(
|\bar q_1 q_1 \rangle
-
|\bar q_2 q_2 \rangle
\right)
,
\\
&&
|{\bf 8}_3 \rangle
=
|\bar q_2 q_1 \rangle
,
\\
&&
|{\bf 8}_4 \rangle
=
|\bar q_1 q_3 \rangle
,
\\
&&
|{\bf 8}_5 \rangle
=
-|\bar q_2 q_3 \rangle
,
\\
&&
|{\bf 8}_6 \rangle
=
|\bar q_3 q_2 \rangle
,
\\
&&
|{\bf 8}_7 \rangle
=
|\bar q_3 q_1 \rangle
,
\\
&&
|{\bf 8}_8 \rangle
=
\frac{1}{\sqrt{6}}
\left(
|\bar q_1 q_1 \rangle
+
|\bar q_2 q_2 \rangle
-2
|\bar q_3 q_3 \rangle
\right)
.
\end{eqnarray*}
Then, the elements of $\hat\rho$ in $\alpha$-represenation
can be related with those in $ij$-representation as
\begin{eqnarray*}
&&
\hat\rho_{{\bf 1},{\bf 1}}
=
|{\bf 1} \rangle\langle {\bf 1}|
=
\frac{1}{3}
\left(
\sum_i
|\bar q_i q_i \rangle
\right)
\left(
\sum_i
\langle\bar q_i q_i|
\right)
=
\frac{1}{3}
\sum_{ij}\hat\rho_{ii,jj}
,
\\
&&
\hat\rho_{{\bf 8_1},{\bf 8_1}}
=
|{\bf 8}_1 \rangle\langle{\bf 8}_1|
=
|\bar q_1 q_2 \rangle\langle \bar q_1 q_2|
=
\hat\rho_{12,12}
,
\\
&&
\hat\rho_{{\bf 8_2},{\bf 8_2}}
=
|{\bf 8}_2 \rangle\langle{\bf 8}_2|
=
\frac12\left(
\hat\rho_{11,11}
+
\hat\rho_{22,22}
-
\hat\rho_{11,22}
-
\hat\rho_{22,11}
\right)
,
\\
&&
\hat\rho_{{\bf 8_3},{\bf 8_3}}
=
|{\bf 8}_3 \rangle\langle{\bf 8}_3|
=
|\bar q_2 q_1 \rangle\langle \bar q_2 q_1|
=
\hat\rho_{21,21}
,
\\
&&
\hat\rho_{{\bf 8_4},{\bf 8_4}}
=
|{\bf 8}_4 \rangle\langle{\bf 8}_4|
=
|\bar q_1 q_3 \rangle\langle \bar q_1 q_3|
=
\hat\rho_{13,13}
,
\\
&&
\hat\rho_{{\bf 8_5},{\bf 8_5}}
=
|{\bf 8}_5 \rangle\langle{\bf 8}_5|
=
|\bar q_2 q_3 \rangle\langle \bar q_2 q_3|
=
\hat\rho_{23,23}
,
\\
&&
\hat\rho_{{\bf 8_6},{\bf 8_6}}
=
|{\bf 8}_6 \rangle\langle{\bf 8}_6|
=
|\bar q_3 q_2 \rangle\langle \bar q_3 q_2|
=
\hat\rho_{32,32}
,
\\
&&
\hat\rho_{{\bf 8_7},{\bf 8_7}}
=
|{\bf 8}_7 \rangle\langle{\bf 8}_7|
=
|\bar q_3 q_1 \rangle\langle \bar q_3 q_1|
=
\hat\rho_{31,31}
,
\\
&&
\hat\rho_{{\bf 8_8},{\bf 8_8}}
=
|{\bf 8}_8 \rangle\langle{\bf 8}_8|
=
\frac16
\left(
\hat\rho_{11,11}
+
\hat\rho_{11,22}
+
\hat\rho_{22,11}
+
\hat\rho_{22,22}
\right.
\\
&&
-
2\hat\rho_{11,33}
-
2\hat\rho_{22,33}
-
2\hat\rho_{33,11}
-
2\hat\rho_{33,22}
+
\left.
4\hat\rho_{33,33}
\right)
.
\\
\end{eqnarray*}

\end{document}